\documentstyle[12pt]{article}
\begin{document}

\titlepage
\begin{flushright}
{\footnotesize{\sf Portsmouth University\\
Relativity and Cosmology Group\\
{\em Preprint} RCG 94/7}}
\end{flushright}
\[ \]
\begin{center}
{\large\bf EXPANDING SPHERICALLY SYMMETRIC MODELS WITHOUT SHEAR}
\end{center}

\vspace{10 mm}

\begin{center}
{\bf S. D. Maharaj$^1$}, {\bf P. G. L. Leach$^{1\dagger *}$}
and {\bf R. Maartens$^{2*}$}
\end{center}
\[ \]
$^1$Department of Mathematics and Applied Mathematics,
University of Natal,
Private Bag X10,
Dalbridge 4014,
South Africa
\[ \]
$^2$School of Mathematical Studies,
University of Portsmouth,
Portsmouth PO1 2EG,
England

\vspace{15 mm}

\noindent
$^\dagger$Centre for Theoretical
and Computational Chemistry, University of Natal, Durban

\vspace{5mm}

\noindent
$^*$Member of Centre for Nonlinear Studies,
University of the Witwatersrand, Johannesburg

\[ \]
\[ \]
\begin{center}
{\bf Abstract}
\end{center} 

The integrability properties of the field equation $L_{xx} = F(x)L^2$
of a spherically symmetric shear--free fluid
are investigated. A first integral, subject to an
integrability condition on $F(x)$, is found, giving a new
class of solutions which contains the
solutions of Stephani (1983) and Srivastava (1987)
as special cases. The
integrability condition on $F(x)$ is reduced to a quadrature
which is expressible in terms of elliptic integrals in general.
There are three classes of solution and in general the solution
of $L_{xx} = F(x)L^2$ can only be written in parametric form.
The case for which $F=F(x)$ can be explicitly given corresponds
to the solution of Stephani (1983).
A Lie
analysis of $L_{xx} = F(x) L^2$ is also
performed. If a constant
$\alpha$ vanishes, then the solutions of Kustaanheimo and Qvist (1948)
and of this paper are regained.
For $\alpha \neq 0$ we reduce the problem to a
simpler, autonomous equation. The applicability of the
Painlev\'e analysis is also briefly considered.

\newpage

\begin{center}
{\bf 1. Introduction}
\end{center}

Spherical symmetry  is often
assumed when seeking exact solutions to
the Einstein field equations for a shear--free matter distribution.
Spherically symmetric solutions may be used as isotropic cosmological
models without the further restriction of homogeneity,
or to model the interior
of expanding or contracting spherical stars (Santos 1985, de Oliveira
{\em et al} 1985). The shear--free condition
substantially simplifies
the field equations and not surprisingly most of the exact solutions
known are not shearing (see Kramer {\em et al} 1980, Krasinski 1989).

For a spherically symmetric, shear--free perfect fluid the
field equations reduce essentially to two nonlinear
partial differential equations.
Various approaches have been followed in seeking solutions to these
equations. McVittie (1933, 1967, 1984)
assumed a functional form of the metric coefficients
which enabled him to solve the field equations to
give a wide class of solutions. The solutions found by
Kustaanheimo and Qvist (1948), Chakravarty {\em et al} (1976), Wyman
(1976) and Stephani (1983) depend on a suitable  choice of a
function of
integration. A more recent approach by
Herrera and Ponce de Leon (1985),
Dyer {\em et al} (1987),
Sussman (1989), Maartens and Maharaj (1990) and Maharaj {\em et al}
(1991) is to suppose that the spacetime
is invariant under a conformal Killing vector, which leads to classes
of solutions.

One of the field equations for a spherically symmetric
shear--free fluid can be reduced to the partial differential
equation $L_{xx} = F(x)L^2$, where $F(x)$ is an arbitrary function
resulting from an integration process.
The purpose of this paper is to investigate integrability properties
of  this equation. In section 2 we give the
field equations for a shear--free matter distribution. A condition
on the function of integration $F(x)$ is found in section 3 that
yields
a first integral of $L_{xx} = F(x) L^2$. We express this condition
as a third order ordinary
differential equation which in general
has solutions in terms of elliptic integrals.
The first integral found does not contain the solution of
Kustaanheimo and Qvist (1948), but does contain those of
Stephani (1983)
and Srivastava (1987)
as special cases. This represents a new class (as far as we are
aware) of solutions, generalising those of Stephani.
In section 4 we perform a Lie analysis on this
equation. If a constant $\alpha$ vanishes, then the solution is
reduced
to a quadrature. One class of the $\alpha =0$ solutions is just the
Kustaanheimo and Qvist solution. The other class
is that found in section 3, thus giving a basis
for the success of the {\em ad hoc} method used in section 3.
For the case $\alpha \neq 0$ we reduce the problem to a
simpler, autonomous equation. In section 5 we consider the possibility
of using the Painlev\'e analysis and discuss results obtained for
$F(x)= x^n$. One consequence of the results is to situate the known
solutions and understand what is special about them.

\begin{center}
{\bf 2. Field equations}
\end{center}

For a spherically symmetric, shear--free, perfect fluid we can
introduce a comoving and isotropic coordinate system
$ x^i = (t,r,\theta ,\phi)$ such that the metric
can be written as
\begin{equation}
ds^2 = - e^{2\nu (t,r)} dt^2 + e^{2\lambda (t,r)} \left[ dr^2 +
r^2(d\theta^2 + \sin^2 \theta d\phi^2)  \right].
\end{equation}
Under these conditions the Einstein field equations reduce to (Kramer
{\em et al} 1980)
\begin{eqnarray}
e^{\nu} &=& \lambda_t e^{-f(t)} \\
e^\lambda ( \lambda_{rr} - \lambda_r^2  - \lambda_r/r)
&=& - {\tilde F}(r),
\end{eqnarray}
where $f(t)$ and ${\tilde F} (r)$ are arbitrary functions
of integration. The energy density $\mu$ and pressure $p$ then assume
the form
\begin{eqnarray}
\mu  &=& 3e^{2f} -e^{-2\lambda} (2\lambda_{rr} + \lambda_r^2  +
4\lambda/r) \\
p &=& {\lambda}^{-1}_{t} e^{-3\lambda} \partial_t [e^\lambda (
\lambda_r^2 + 2\lambda_r/r) - e^{3\lambda + 2f} )].
\end{eqnarray}
To find an exact solution of the field equations (2)--(3)
the functions $f(t)$ and ${\tilde F} (r)$ need to be specified and
equation (3) has to be solved for $\lambda$. Then the
dynamical quantities $\mu$ and $p$
can be computed from (4) and (5).

The transformation (Kustaanheimo and Qvist 1948)
\[
x=r^2, \,\,\,\, L(t,x) = e^{-\lambda}, \,\,\,\,
{\tilde F}(x) = F/4r^2
\]
reduces equation (3) to
\begin{equation}
L_{xx} = F(x) L^2.
\end{equation}
We will be concerned with this form of the field equation in
subsequent sections. In spite of the tremendous interest generated by
spherically symmetric shear--free perfect fluids the solution of the
field equations in complete generality is known only for
(Kustaanheimo and Qvist 1948)
$$
F(x) = {(ax^2 + bx +c)}^{-5/2} \eqno{(7a)}
$$
and (Stephani 1983)
$$
F(x) = x^{-15/7} \,\,\,\, \mbox{or} \,\,\,\, x^{-20/7}.
\eqno{(7b)}
$$
Incidentally subclasses of the
McVittie (1984) class of metrics also correspond
to these forms of $F(x)$.
We extend this to a new class of solutions in section 3.

As pointed out in Kramer {\em et al} (1980), the invariance of (6)
under coordinate transformations that preserve the form of (1),
implies that $F$ transforms as
$$
F(x) \rightarrow \tilde F ( \tilde x ) = {\tilde x}^{-5}
F( {\tilde x}^{-1} ) \eqno{(8a)}
$$
under
$$
x \rightarrow \tilde x = x^{-1}, \,\,\,\, L \rightarrow \tilde L
= x^{-1} L
\eqno{(8b)}
$$
(Note the erroneous ${\tilde x}^5$, instead of ${\tilde x}^{-5}$,
given on p 168 of Kramer {\em et al}.) This follows since
\[
\frac{\partial^2L}{\partial x^2} - F(x) L^2 = {\tilde x}^3 \left[
\frac{\partial^2 \tilde L}{\partial{\tilde x}^2} - {\tilde x}^{-5}F(x)
{\tilde L}^2 \right].
\]
Thus a solution for a given $F(x)$ is equivalent to a solution
with $F(x)$ replaced by $x^{-5} F(x^{-1})$. Note that (7$a$) is
form invariant under (8), while the pair in (7$b$)
are equivalent.

Note also that the solutions of (6) with $F= (ax+b)^n$ are trivially
related to the solutions with $F=x^n$. However, the two
types of solution are not equivalent, since the metric is not
invariant
under $x \rightarrow ax+b$ (Stephani 1983, Srivastava 1987).

\begin{center}
{\bf 3. A first integral of $L_{xx} = F(x) L^2$ }
\end{center}

Here we extend the technique of Srivastava (1987) to obtain
a first integral of (6) without choosing an explicit functional
form for $F(x)$. The first integral found is subject to an integral
equation in $F(x)$ which can also be expressed as a nonlinear third
order ordinary differential equation.

A number of choices of $F(x)$ is known that reduces the order
of the field equation (6) (Srivastava 1987, Stephani 1983).
Rather than choose $F(x)$ we seek a general condition
on $F(x)$ that reduces  (6) to a first order differential equation.
The integral of $F(x)L^2$ can be expressed as
\setcounter{equation}{8}
\begin{equation}
\int F(x) L^2 dx = F_I L^2 - 2 \int F_I L L_x dx,
\end{equation}
where for convenience we have used the notation
\[
F_I = \int F(x)dx.
\]
Integrating $F_I LL_x$ by parts and using (6) gives the result
\[
\int F_I L L_x dx = F_{II} LL_x - \int F_{II} L_x^2 dx  -
\int FF_{II} L^3 dx
\]
On further evaluating the integrals on the right hand side of this
equation and substituting in (9) we obtain
\begin{eqnarray}
\int F(x) L^2 dx &=& F_I L^2 -2 F_{II} LL_x + 2 F_{III} L^2_x
+2 (FF_{II})_I L^3 \nonumber \\
& & -2 \left[ \int \left[
2FF_{III} + 3(FF_{II})_I \right] L^2 L_xdx \right]
\end{eqnarray}
We observe that the integral on the right hand side of (10)
can be evaluated if $2FF_{III} + 3(FF_{II})_I $ is
a constant.
Then equations (6) and  (10) yield the result
\begin{equation}
\psi_0 (t) = -L_x + F_I L^2 -2 F_{II} LL_x + 2 F_{III} L^2_x
+ 2 \left[  (FF_{II})_I - \mbox{$\frac{1}{3}$} K_0 \right] L^3
\end{equation}
when we impose the condition
\begin{equation}
2FF_{III} + 3 (FF_{II})_I = K_0,
\end{equation}
where $\psi_0(t)$ is an arbitrary function of integration and $K_0$ is
a constant. Thus we have established that {\em (11) is a first
integral
of the field equation (6) provided that condition (12) is satisfied.}

We transform the integral equation (12) into an ordinary
differential equation
which is easier to work with. Differentiating equation (12) gives
\begin{equation}
2F_x F_{III} + 5F F_{II} = 0
\end{equation}
The form of (13) suggests the transformation
\begin{equation}
{\cal F} \equiv  F_{III}.
 \end{equation}
Substitution of  (14) into
(13) gives
\[
2{\cal F} {\cal F}_{xxxx} + 5 {\cal F}_{x}{\cal F}_{xxx} = 0
\]
with solution
\begin{equation}
{\cal F}_{xxx} = K_1 {\cal F}^{-5/2},
\end{equation}
where $K_1$ is a constant of integration. Note that $K_1 =0$
implies $F=0$ by (15) and (14).
We have established that the third order
ordinary equation (15) with definition (14) is equivalent to the
integrability condition (12). A solution of (15) gives ${\cal F}$ and
then
$F(x)$ is found from (14). Repeated integration of (15) yields the
result
\begin{equation}
{\cal F}^{-1} =  K_4 + K_3 \left( {\int } {\cal F}^{-3/2} dx\right)  +
K_2 {\left( {\int} {\cal F}^{-3/2} dx  \right)}^2 -
\frac{1}{6} K_1 {\left(
{\int} {\cal F}^{-3/2} dx \right)}^3,
\end{equation}
where $K_2,K_3,K_4$ are constants of integration.
As this integration is nontrivial we provide details in the appendix.
At this point it is
convenient to let
\[
u = {\int} {\cal F}^{-3/2} dx
\]
so that
\begin{equation}
u_x = {({\cal F}^{-1})}^{3/2}.
\end{equation}
Substitution of (16) into (17)  gives the result
\begin{equation}
x - x_0 = \int \frac{du}{{(K_4 + K_3 u + K_2 u^2 - \frac{1}{6}
K_1 u^3)}^{3/2}},
\end{equation}
where $x_0$ is constant. Thus we have reduced the differential
equation
(15) to the quadrature (18) which can be evaluated in terms of
elliptic
integrals in general (Gradshteyn and Ryzhik 1994).
In summary: {\em the first integral (11), with $F$ given by (18),
represents a class of solutions of (6).}

It would be of interest to identify those cases where we can perform
the
inversion $u=u(x)$ and explicitly write $F=F(x)$:
Once the integral in (18) is evaluated we invert
to obtain $u = u(x)$ and then (17) gives ${\cal F} =
u_{x}^{-2/3}$. Finally the function $F(x)$ is found from
$F ={\cal F}_{xxx}$. This $F$ then satisfies the integrability
condition (12) for the existence of the first integral
(11).

Hence to find solutions $F(x)$ in closed form,
satisfying the integrability condition (12), we need to
evaluate the integral (18). It is convenient to express our
solution to (12) in  the parametric form
\begin{eqnarray*}
F(x) &=& {\cal F}_{xxx} \\
u_x &=& {\cal F}^{-3/2} = \left[ g^\prime (u) \right]^{-1} \\
x- x_0 &=& g(u),
\end{eqnarray*}
where for convenience we have set
\[
g(u) =
\int \frac{du}{{(K_4 + K_3 u + K_2 u^2 - \frac{1}{6}
K_1 u^3)}^{3/2}}.
\]
Three cases arise in the above solution depending
on the nature of the
factors of the polynomial
$K_4 + K_3 u + K_2 u^2 - \frac{1}{6}K_1 u^3$:\\

\noindent
{\em Case I: Three repeated factors}

If there are three repeated factors then we can write
\[
K_4 + K_3 u + K_2 u^2 - \frac{1}{6}K_1 u^3
= (a+bu)^3 \qquad b \neq 0.
\]
In this case (18) can be integrated immediately in terms of
elementary functions and we obtain
\begin{eqnarray*}
g(u) &=& -\frac{2}{7b} (a+bu)^{-7/2} \\
F(x) &=& \frac{48}{343} \left(- \frac{7b}{2} \right)^{6/7}
(x-x_0)^{-15/7}.
\end{eqnarray*}
Thus in this case it is possible to invert the integral (18)
and to write $u =u(x)$.
After reparametrisation we can write
\[
F(x) = x^{-15/7}
\]
which is the same as the case reported by Stephani (1983).
We recall that this
also leads to the $x^{-20/7}$ solution as noted in section 2.\\

\noindent
{\em Case II: Two repeated factors}

With two repeated factors we can write
\[
K_4 + K_3 u + K_2 u^2 - \frac{1}{6}K_1 u^3
= (a+bu)(u+c)^2   \qquad b \neq 0
\]
The function $g(u)$ in (18) may be easily integrated for two
repeated factors so that
\begin{eqnarray*}
g(u) &=&  \left(- \frac{1}{ 2(a-bc)u^2} + \frac{5b}{4 (a-bc)^2 u} +
\frac{15b^2}{4 (a-bc)^3 } \right) \frac{1}{\sqrt{a-bc +bu}} + \\
& &
\frac{15b^2}{8 (a-bc)^3 } \int \frac{du}{u \sqrt{a-bc +bu}},
\end{eqnarray*}
where the integral on the right side is expressible
in terms of elementary functions.  (The precise form depends upon the
sign of
$a - bc$.  See Gradshteyn and Ryzhik (1994) 2.246 and 2.248.9.)
It is clear from the above form for $g(u)$ that it will not be
possible
to perform the inversion to obtain $u=u(x)$. (This is possible only
when $b=0$ which is not permitted.)
Consequently the solution
can only be given parametrically in this case.\\

\noindent
{\em Case III: No repeated factors}

If there are no repeated factors, we write
\[
K_4 + K_3 u + K_2 u^2 - \frac{1}{6}K_1 u^3
= d(a-u)(b-u)(c-u)  \qquad d \neq 0.
\]
In this case the solution of (18) is given in terms of elliptic
integrals. We obtain
\begin{eqnarray*}
g(u) &=& d^{-3/2} \frac{2[c(a-c) +b(a-b)-u(2a-c-b)]}{(a-b)(a-c)(b-c)^2
\sqrt{(a-u)(b-u)(c-u)}} + \\
& & d^{-3/2} \frac{2}{(a-b)^2(b-c)^2\sqrt{(a-c)^3}} \times \\
& &
\times \left[ (b-c)(a+b-2c) F(\alpha, p)
- 2(c^2 +a^2 +b^2 -ab -ac -bc) E(\alpha, p) \right],
\end{eqnarray*}
where we have set
\[
\alpha =\mbox{arcsin} \sqrt{\frac{a-c}{a-u}} \qquad
 p= \sqrt{\frac{a-b}{a-c}}.
\]
In the above $F(\alpha,p)$ is the elliptic
integral of the first kind and $E(\alpha,p)$
is the elliptic integral of the second kind. By inspection of
$g(u)$ we observe that it is not possible to invert and obtain
$u=u(x)$. Therefore, as in {\em Case II},
the solution may be expressed
only in parametric form when there are no repeated factors.

We have established that for our class of first integrals
the form of $F$ may be given in parametric form in general.
It is only in {\em Case I} that inversion is possible and we can write
$F=F(x)$ explicitly. This case corresponds to
$F=x^{-15/7}$ which was first identified by Stephani (1983).

It is interesting to observe that the solution of Kustaanheimo and
Qvist (1948) is not contained in the first integral (11).
To prove this
we simply substitute (7$a$) into (12) and see that (12)
is not satisfied.
Thus the first integral (11) cannot admit
the Kustaanheimo and Qvist (1948) solution with (7$a$). The first
integral (11), with $F$ given by the quadrature (18),
 represents a new class of spherically symmetric solutions. Most
papers have dealt with special cases of (7$a$) (Kramer {\em et al}
1980).

Note that  the first integral (11) was obtained
without specifying the function $F(x)$.
However, in our solution $F(x)$ is constrained by the
integral equation (12). Thus any $F(x)$ satisfying (12) will  yield
a first integral of the form (11). The solutions of Srivastava (1987)
and
Stephani (1983) can be regained as special cases of (11). With the
choice
$F (x) = {(ax+b)}^n$ where $n= -15/7$ we find that (12)
is satisfied with  $K_0=0$.
Then (11) becomes
\begin{eqnarray}
\phi_0 (t) &\equiv&     (1/6) \psi_0 (t) \nonumber \\
&=& - 6L_x - (21/4a) {(ax+b)}^{-8/7} L^2 -
(3/2) {(7/a)}^2 {(ax+b)}^{-1/7} LL_x  \nonumber \\
& & + (1/4){(7/a)}^3 {(ax+b)}^{6/7}
L_x^2 - (1/6) {(7/a)}^3 {(ax+b)}^{-9/7} L^3.
\end{eqnarray}
The first integral (19) was reported by Srivastava (1987).
Now with $a=1$ and $b=0$ equation (19) reduces to
\begin{eqnarray}
\varphi_0 (t) &\equiv& (1/6) \psi_0 (t) \nonumber \\
&=& - 6L_x - (21/4) x^{-8/7} L^2 -
(3/2) 7^2 x^{-1/7} LL_x + (1/4)7^3 x^{6/7}
L_x^2 \nonumber \\
& &- (1/6) 7^3 x^{-9/7} L^3.
\end{eqnarray}
The first integral (20), where $F(x)$ is given by (8), was found by
Stephani (1983). Note that in
(19) and (20) the functions $\phi_0 (t)$ and $\varphi_0 (t)$ have been
introduced to
ease comparison with the corresponding expressions given by Srivastava
(1987) and Stephani (1983) respectively. Hence the first integral (11)
represents a class of solutions which may be interpreted
as a generalisation of the Stephani/Srivastava solution. In fact,
the $F$ of the form $(ax+b)^n$ only satisfies (12) if $n=-15/7$
(which can be mapped to the case $n=-20/7$ by (8)). This is a special
solution of (12). The general solution, containing the generalisation
of Stephani and Srivastava, is given by (18).
The case $n=-15/7$ is characterised by the fact that it is the
only case for which we can write $F=F(x)$ explicitly; all
other solutions may only be given parametrically.

\begin{center}
{\bf 4. Lie Analysis}
\end{center}

In section 3 we obtained the first integral (11) of (6) in an
{\em ad hoc} fashion. Here we consider a widely applied technique
to reduce (6) to a first order  differential equation. A systematic
method to determine whether  a second order ordinary differential
equation can be solved by quadratures is that of Lie (1912). On
performing the Lie analysis (Leach 1981, Leach {\em et al} 1992)
we find that (6) is invariant under the infinitesimal transformation
\begin{equation}
{\bf G} = A(x) \frac{\partial}{\partial x} +
[B(x) L + C(x)] \frac{\partial}{\partial L}
\end{equation}
provided that
\begin{eqnarray}
F(x) &=& MA^{-5/2} \exp \left( \frac{\alpha}{2} \int \frac{dx}{A}
\right) \\
A_{xxx} &=& 2 MCA^{-5/2} \exp \left( \frac{\alpha}{2} \int\frac{dx}{A}
\right) \\
B(x) &=& (A_x -\alpha )/2 \nonumber \\
C(x) &=& C_0 + C_1 x \nonumber,
\end{eqnarray}
where $\alpha , M, C_0$ and $C_1$ are constants. A solution of (23)
gives
$F(x)$ by (22) which reduces (6) to a first order  differential
equation. Note that the existence of the symmetry (21) permits a
reduction to first order. If two symmetries ${\bf G}$ are known,
the second order equation (6) may be reduced to quadratures.

For convenience we divide the integration of (23) into a number of
cases:\\

\noindent
{\em Case I}: $\alpha =0 \qquad C=0$

This is the simplest case. Then (22) yields
\[
F(x) = M{(ax^2 + bx +c)}^{-5/2}
\]
which is equivalent to the Kustaanheimo and Qvist (1948)
class of solutions.\\

\noindent
{\em Case II}: $\alpha =0 \qquad C_0 \neq 0 = C_1$

In this case (23) reduces to the differential equation
\[
A_{xxx} = K_1 A^{-5/2},
\]
where $K_1 =2MC_0$. This is the same as equation (15) considered
in section 3 and the results of that section become
applicable. Thus we have established that  the first integral
(11) is a special class of solutions admitted by the general
Lie method.\\

\noindent
{\em Case III}: $\alpha =0  \qquad C_1\neq 0$

We now consider the case $C_1 \neq 0$.
 We begin by observing that (23) is simplified
by a change of the independent variable $x$ which transforms away the
linear term $C(x) = C_0 + C_1x$. If we introduce the canonical
variables
($C_1 \neq 0$)
\[
Y = A/C^2 , \,\,\,\, X = -1/C,
\]
(23) is transformed to
\begin{equation}
Y^{\prime\prime\prime} = M_1 Y^{-5/2}
\end{equation}
where $M_1 = 2MC_1^{-3}$,  and we have used the notation
$^\prime :=d/dX$.  We provide an alternate solution to (24)
than that presented in section 3.
Equation (24) can be reduced to an
equation involving only $Y^\prime$ if we introduce the variable
\begin{equation}
s := \int Y^{-3/2} dX
\end{equation}
Then (24) and (25) imply
\[
YY^\prime s^\prime  = -M_3
-2M_2s + \mbox{$\frac{1}{2}$}M_1s^2,
\]
where $M_2$ and $M_3$ are constants. On multiplying this equation by
$s^\prime$ and integrating we obtain
\begin{equation}
Y^2s^{\prime 2}
 = M_4 + M_3s + M_2s^2 -\mbox{$\frac{1}{6}$}M_1s^3,
\end{equation}
where $M_4$ is constant.
Equation  (26) is reduced to the quadrature
\begin{equation}
x - x_1 = \int \frac{ds}{ {(M_4 + M_3s + M_2s^2
-\frac{1}{6}M_1s^3)}^{3/2} },
\end{equation}
where $x_1$ is a constant of integration. As mentioned in section 3
integrals of the form (27) can be evaluated in terms of elliptic
integrals in general in which case the solution
will be given parametrically.
However, in some cases
after having solved (27) we then invert to obtain
$s(X)$ and finally we have $A = C^2Y = C^2{(s^\prime)}^{2/3}$.\\

\noindent
{\em Case IV}: $\alpha \neq 0 \qquad C =0$

It remains to consider the highly complicated case of $\alpha \neq
0$.
Apart from the subcase $C=0$,
we have not been able to integrate (23) which is hardly
surprising considering its highly nonlinear nature.
The special case is  $\alpha \neq 0 =C$. This appears to give a
generalisation of
Kustaanheimo and Qvist since (22) and (23) imply
\begin{eqnarray*}
A &=& ax^2 +bx +c \\
F &=& MA^{-5/2} \exp \left( \int \frac{\alpha}{2A} dx \right).
\end{eqnarray*}
Now let
\begin{eqnarray*}
\cal Y &=& L A^{-1/2} \exp \left( \int \frac{\alpha}{2A} dx \right) \\
\cal X &=& \int \frac{dx}{A}.
\end{eqnarray*}
Then (6) is autonomised:
\begin{equation}
{\cal Y}^{\prime\prime} -\alpha {\cal Y}^\prime +
\left( \frac{\alpha^2}{4} + ac  - \frac{1}{4}b^2 \right) {\cal Y}
+ M {\cal Y}^2 = 0.
\end{equation}
Now, if $\Delta \equiv b^2 -4ac > 0$ and $\alpha =\pm 5\sqrt{\Delta}$,
then (28) can be transformed to
\begin{equation}
\frac{d^2\eta}{d\xi^2} + \frac{M}{\Delta} \eta^2 =0,
\end{equation}
where
\begin{eqnarray*}
\eta &=& e^{-2\alpha {\cal X} /5} {\cal Y} \\
\xi &=& e^{\alpha {\cal X} /5}.
\end{eqnarray*}
Then (29) is easily integrated to yield
\[
\xi - \xi_0 = \int \frac{d\eta}{ [C - (2M/3\Delta)\eta^3]^{1/2} }.
\]
This seems to be a new solution, but
\[
\Delta > 0 \Rightarrow A =ax^2 +bx+c = a(x-\mu )(x-\nu )
\]
with
$\mu - \nu = \sqrt{\Delta}/a$ ($\neq 0$). Then
\[
\int \frac{dx}{A}  = \frac{1}{\sqrt{\Delta}} \ln \left(
\frac{x-\mu}{x-\nu} \right)
\]
implies
\[
F = \left\{ \begin{array}{ll}
    (Ka^{-5/2}) (x-\mu)^{-5} & \alpha = 5\sqrt{\Delta} \\
    (Ka^{-5/2}) (x-\nu)^{-5} & \alpha = -5\sqrt{\Delta}
\end{array} \right.
\]
which remarkably is a special case of Kustaanheimo and Qvist!
Thus even with $\alpha \neq 0 = C$ we do not obtain any new solutions.

We note that  if instead of assuming that $\alpha$
has the value that enables us to transform
(28) to (29) other functions $F$ can be obtained. For example
if we put $a=b=0$ in $A$,
\[
F(x) = Mc^{-5/2} e^{\alpha x/(2c)}
\]
which is equivalent to the Stephani (1983) form
\[
F(x) = e^x
\]
after reparametrisation. When $A= a(x-P)(x-Q)$,
\[
F(x) = Ma^{-5/2} \left(x-P\right)^{\frac{\alpha}{2a(P-Q)}-\frac52}
\left(x-Q\right)^{-\frac{\alpha}{2a(P-Q)} -\frac52}
\]
which reduces to to the Stephani (1983) form
\[
F(x) = (x+\alpha )^n (x+\beta )^{-n-5}
\]
on the substitution $\frac{\alpha}{2a(P-Q)} = n + \frac52$.
If $A = a (x-P)^2$ then
\[
F(x) = M a^{-5/2} \left(x-P \right)^{-5}
e^{ -\frac{\alpha}{2a} \frac{1}{x-P}}
\]
which via (8a) is related to $F(x) = e^x$.\\

\noindent
{\em Case V}: $\alpha \neq 0 \qquad C_1 \neq 0$

This is the most difficult case to analyse.
When $C_1 \neq 0$, we are in a
position to express (23) as a simpler autonomous equation after a
change
of variable. Introduction of the variable
\begin{equation}
{\cal A} = \exp \left( \frac{\alpha}{5} \int \frac{dx}{A} \right)
\end{equation}
transforms (23) to the differential equation
\begin{equation}
{ \left( \frac{{\cal A}}{{\cal A}_x} \right) }_{xxx} =
2{(5/\alpha )}^{7/2}MC {\cal A}^{-5/2}_{x}.
\end{equation}
For $C_1\neq 0$ without any loss of generality we can
write (31) as
\begin{equation}
{ \left( \frac{{\cal A}}{{\cal A}_x} \right) }_{xxx} = x{\cal
A}^{-5/2}_{x},
\end{equation}
where we have set $C_0 = 0$ and $2{(5/\alpha )}^{7/2}C_1 = 1$.
Essentially we have rescaled the $x$ variable by ${\tilde x}
\longrightarrow \beta x + \gamma$. Expansion of the left hand side of
(32) results in
\begin{equation}
{\cal A}{\cal A}^{2}_{xx}{\cal A}_{xxxx} + 2{\cal A}^{3}_{x}{\cal
A}_{xxx} - 6{\cal A}{\cal A}_{x}{\cal A}_{xx}{\cal A}_{xxx}
- 3{\cal A}^{2}_{x}{\cal A}^{2}_{xx} + 6{\cal A}{\cal A}^{3}_{xx} + x
{\cal A}^{13/2}_{x} = 0.
\end{equation}
The fourth order equation (33) is reduced to autonomous form if we
introduce the new independent variable
\[
r = -\frac{1}{x}.
\]
With this transformation (33) assumes the autonomous form
\begin{equation}
{\cal A}{\cal A}^{2}_{rr}{\cal A}_{rrrr} + 2{\cal A}^{3}_{r}{\cal
A}_{rrr} - 6{\cal A}{\cal A}_{r}{\cal A}_{rr}{\cal A}_{rrr}
- 3{\cal A}^{2}_{r}{\cal A}^{2}_{rr} + 6{\cal A}{\cal A}^{3}_{rr} +
{\cal A}^{13/2}_{r} = 0.
\end{equation}
The order of (34) is reduced by the substitution
\[
{\cal A} = e^z , \qquad {\cal A}_r = 4Z^2e^{-4z}
\]
to
\begin{equation}
ZZ_{zzz} - Z_zZ_{zz} -5ZZ_{zz} + 5Z_{z}^2 +Z^3 = 0.
\end{equation}
Thus we have expressed equation (23) with $\alpha \neq 0$ in the
simpler
form (35). A solution of (35) gives ${\cal A}$ and $A$ is then found
from (30).

\begin{center}
{\bf 5. Discussion}
\end{center}

We have considered the integrability of the equation $L_{xx} =
F(x)L^2$.
A first integral was found in section 3 that contains the special
cases
of Stephani (1983) and Srivastava (1987)
studied previously. The solution is given parametrically
and there are three classes of solution. In one case
we can perform the inversion and write $F=F(x)$
explicitly.  This corresponds to $F= x^{-15/7}$ obtained by
Stephani.
In section 4 the Lie analysis of this equation
resulted in the equation (23). For $\alpha = 0$ equation (23) has
solutions in terms of elliptic integrals. In the more complicated
case
of $\alpha \neq 0$ we reduced (23) to the simpler autonomous equation
(35).

Considering the significance of spherically symmetric models in
general
relativity the equation $L_{xx} = F(x)L^2$ deserves further attention.
Another approach in the investigation of the integrability of this
equation is the Painlev\'e analysis.
In the Painlev\'e analysis one performs a singularity analysis
of a differential equation. If the only movable singularities are
poles, the equation is said to possess the Painlev\'e property
and it is conjectured that such equations can be solved
analytically (Ramani {\em et al} 1989).

Omitting the details we illustrate this
approach with an example . With $F(x) = x^n$ Stephani (1983) obtained
an
ansatz that reduced
\begin{equation}
L_{xx} = x^nL^2
\end{equation}
to a first order equation.
We performed the Painlev\'e analysis on equation (36) after
transforming
it to autonomous form by the transformation which converts the
appropriate form of $\bf G$ in (21) to $\partial /\partial X$.
The autonomous form of (36) is
\begin{equation}
Y^{\prime\prime} - (2n+5) Y^\prime + (n+2) (n+3)Y - Y^2 = 0
\end{equation}
and satisfies the Painlev\'e property for the cases
$n=-5/2, 0, -5, -15/7$ and $-20/7$. These are precisely
the cases of known solutions, i.e. Kustannheimo and Qvist (1948)
($n=-5/2, 0, -5$) and Stephani ($n =-15/7, -20/7$).

\newpage

\begin{center}
{\bf Appendix}
\end{center}

In this appendix we indicate how the third order equation
(15) may be integrated to yield the solution (16).
The nonlinear equation
\[
{\cal F}_{xxx} = K_1 {\cal F}^{-5/2}
\]
may be written as
\[
\left( {\cal F} {\cal F}_{xx} \right)_x - \frac{1}{2}
\left( {\cal F}_{x}{}^2 \right)_x = K_1 {\cal F}^{-3/2}
\]
This gives
\[
 {\cal F} {\cal F}_{xx}  - \frac{1}{2}
 {\cal F}_{x}{}^2  = -2K_2 + K_1
{\left( {\int} {\cal F}^{-3/2} dx\right)}
\]
where $-2K_2$ is a constant. This second order equation may be written
in the form
\[
2 \left({\cal F}^{1/2} \right)_{xx}
 = -2K_2 {\cal F}^{-3/2} + K_1 {\cal F}^{-3/2}
{\left( {\int}{\cal F}^{-3/2}dx\right)}
\]
On integration we obtain
\[
2 \left({\cal F}^{1/2} \right)_x
 = -K_3 -2K_2 {\left( {\int} {\cal F}^{-3/2} dx\right)}
+ \frac{1}{2} K_1
{\left( {\int}{\cal F}^{-3/2}dx\right)}^2
\]
where
$-K_3$ is a constant.  It is convenient to rewrite this
first order equation in the form
\[
-{\cal F}^{-2} {\cal F}_x =  K_3  {\cal
F}^{-3/2}  +
2 K_2 {\cal F}^{-3/2} {\left( {\int} {\cal F}^{-3/2}dx  \right)} -
\frac{1}{2} K_1 {\cal F}^{-3/2} {\left(
{\int} {\cal F}^{-3/2} dx \right)}^2
\]
A final integration generates the solution
\[
{\cal F}^{-1} =  K_4 + K_3 \left( {\int} {\cal F}^{-3/2}dx \right)  +
K_2 {\left( {\int}{\cal F}^{-3/2} dx \right)}^2 -
\frac{1}{6} K_1 {\left(
{\int} {\cal F}^{-3/2}  dx \right)}^3
\]
where $K_4$ is a constant.

\vspace{1cm}

\noindent
{\bf Acknowledgements:} SDM thanks Dave Matravers and the
University of Portsmouth for hospitality and support where this work
was completed. RM and SDM thank the FRD for a research grant.


\newpage

\begin{thebibliography}{99}
\bibitem{} Chakravarty N, Dutta Choudhury S B and Banerjee A
1976 {\em Austral. J. Phys.} {\bf 29} 113
\bibitem{} de Oliveira A K G, Santos N O and Kolassis C A 1985
{\em Mon. Not. R. Astr. Soc.} {\bf 216} 1001
\bibitem{} Dyer C C, McVittie G C and Oates L M 1987
{\em Gen. Relat. Grav.} {\bf 19} 887
\bibitem{} Gradshteyn I S and Ryzhik I M 1994 {\em
Table of Integrals, Series and Products}
(New York: Academic Press)
\bibitem{} Herrera L and Ponce de Leon J 1985 {\em J. Math.
Phys.} {\bf 26} 778, 2018, 2847
\bibitem{} Kramer D, Stephani H, MacCallum M A H and Herlt E 1980
{\em Exact Solutions of Einstein's Field Equations} (Cambridge:
Cambridge University Press)
\bibitem{} Krasinski A 1989 {\em J. Math. Phys.} {\bf 30} 438
\bibitem{} Kustaanheimo P and Qvist B 1948 {\em Soc. Sci.
Fennica, Commentationes Physico--Mathematicae} {\bf XIII} 16
\bibitem{} Leach P G L 1981 {\em J. Math. Phys.} {\bf 22} 465
\bibitem{} Leach P G L, Maartens R and Maharaj S D 1992 {\em Int. J.
Nonlin. Mech.} {\bf 27 } 575
\bibitem{} Lie S 1912 {\em Vorlesungen \"{u}ber
Differentialgleichungen} (Leipzig und Berlin: Teubner)
\bibitem{} Maartens R and Maharaj M S 1990 {\em J. Math.
Phys.} {\bf 31}
\bibitem{} Maharaj S D, Leach P G L and Maartens R 1991 {\em Gen.
Relat. Grav.} {\bf 23} 261
\bibitem{} McVittie G C 1933 {\em Mon. Not. R. Astr. Soc.}
{\bf 93} 325
\bibitem{} McVittie G C 1967 {\em Ann. Inst. Henri Poincar\'e}
{\bf 6} 1
\bibitem{} McVittie G C 1984 {\em Ann. Inst. Henri Poincar\'e}
{\bf 40} 325
\bibitem{} Ramani A, Grammaticos B and Bountis T 1989
{\em Phys. Rep.} {\bf 180} 159
\bibitem{} Santos N O 1985 {\em Mon. Not. R. Astr. Soc.}
{\bf 216} 403
\bibitem{} Srivastava D C 1987 {\em Class. Quantum Grav.} {\bf 4} 1093
\bibitem{} Stephani H 1983 {\em J. Phys. A: Math. Gen.}
{\bf 16} 3529
\bibitem{} Sussman R A 1989 {\em Gen. Relat. Grav.} {\bf 12} 1281
\bibitem{} Wyman M 1976 {\em Can. Math. Bull.} {\bf 19} 343
\end{thebibliography}
\end{document}